\documentclass[a4paper,10pt]{article}
\usepackage{graphicx}
\usepackage{comment}
\usepackage{amssymb}
\usepackage{amsmath}
\usepackage{nicefrac}
\usepackage{graphicx}
\usepackage{dcolumn}
\usepackage{bm}
\usepackage{comment}
\usepackage{multirow}
\usepackage{cite}
\usepackage{mathrsfs}
\usepackage{url}

\begin{document}

\huge

\begin{center}
Checking the reliability of opacity databases
\end{center}

\vspace{0.5cm}

\large

\begin{center}
Jean-Christophe Pain$^{a,b,}$\footnote{jean-christophe.pain@cea.fr} and Patricia Croset$^a$
\end{center}

\normalsize

\begin{center}
\it $^a$CEA, DAM, DIF, F-91297 Arpajon, France\\
\it $^b$Universit\'e Paris-Saclay, CEA, Laboratoire Mati\`ere en Conditions Extr\^emes,\\
\it 91680 Bruy\`eres-le-Ch\^atel, France
\end{center}

\abstract{Mathematical inequalities, combined with atomic-physics sum rules, enable one to derive lower and upper bounds for the Rosseland and/or Planck mean opacities. The resulting constraints must be satisfied, either for pure elements or mixtures. The intriguing law of anomalous numbers, also named Benford's law, is of great interest to detect errors in line-strength collections required for fine-structure calculations. Testing regularities may reveal hidden properties, such as the fractal nature of complex atomic spectra. The aforementioned constraints can also be useful to assess the reliability of experimental measurements. Finally, we recall that it is important to quantify the uncertainties due to interpolations in density-temperature opacity (or more generally atomic-data) tables, and that convergence studies are of course unavoidable in order to address the issue of completeness in terms of levels, configurations or superconfigurations, which is a cornerstone of opacity calculations.}

\section{Introduction}

The radiative opacity (or mass absorption coefficient) is a key ingredient of stellar models \cite{Armstrong1972,Rybicki1985,Huebner2016,Michaud2016,Pain2018b}. In the complex multi-physics simulations of stellar structure and evolution, the opacities are usually not computed ``in line'' at each time step and at each radial mesh - the numerical cost would be too high - but taken from pre-computed tables. This is possible under the assumption of local thermodynamic equilibrium (LTE), where the only parameters are density and temperature (electronic and ionic). Out of equilibrium, the radiation field, required for the collisional-radiative model, must be determined at each time step and each region of the plasma. This occurs for instance in the simulation of laser experiments dedicated to inertial confinement fusion studies. However, even in that case, opacity tables are often required, combined with the use of effective temperatures \cite{Ralchenko,Bauche2015}.

It is therefore of primary importance to check the reliability of opacity tables. In course of the Orion project \cite{Dyson2002}, Dyson noticed that quantum mechanics enables one to obtain bounds on opacities. This led Bernstein and him to write a report \cite{Bernstein1959}, which was published in the open literature only in 2003 \cite{Bernstein2003}. The bound, obtained using the mathematical Schwarz inequality and the Thomas-Reiche-Kuhn oscillator-strength sum rule, was cited by Armstrong \cite{Armstrong1962}, who proposed, using the results of Refs. \cite{Bernstein1959,Bernstein2003}, an inequality involving both the Planck and Rosseland mean opacities. 

Starting from the oscillator-strength sum rule, Imshennik \emph{et al.} derived an integral relation which must be satisfied by the bound-electron radiation absorption coefficient when the distribution of ions with respect to degree of ionization and excitation state is arbitrary. Making use of such a relation, the authors formulated and solved a variational problem which, in LTE conditions, yields the smallest possible value of the Rosseland mean free path, \emph{i.e.} the largest possible value of the Rosseland opacity \cite{Imshennik1986}.

In a quite similar vein, Molodtsov \emph{et al.} constructed a complete set of estimates for the maximal Rosseland mean opacity for a LTE plasma, on the basis of quantum-mechanical sum rules of the kind
\begin{equation}\label{molo}
\int_0^{\infty}\kappa(\nu)~ \nu^k~d\nu,
\end{equation}
where $\nu$ is the photon frequency and $\kappa$ the radiative opacity. The case $k=0$ is a direct consequence of the Thomas-Reiche-Kuhn sum rule and is equal to the number of electrons in the atomic system. The case $k=-1$ can be expressed through the mean square radius of the atom in the ground state, the $k=+1$ sum through the mean square momentum of the electron in the ground state and $k=+2$ in terms of the density of the electrons at the nucleus \cite{Molodtsov1993}. 

Using mathematical inequalities (such as the Schwarz, H\"older or Milne ones), we first discuss the derivation and relevance of known and new bounds, either for pure elements or mixtures. Then, we recall that the intriguing law of anomalous numbers, also named Benford's law, is of great interest to detect errors in line-strength collections that are required to perform fine-structure calculations. In the same spirit, we emphasize the fact that testing regularities, such as the Learner rule, can reveal hidden (in the present instance fractal) properties. Finally, we insist on the need to quantify the uncertainties due to interpolations in density-temperature opacity (or more generally atomic-data) tables and illustrate the importance of convergence studies. This concerns for instance the number of levels, configurations and / or superconfigurations included in the calculation. In the present work, all our opacities are, for simplicity, computed in the framework of the Super Transition Arrays (STA) approach. We therefore consider the convergence with respect to the number of superconfigurations. 

The paper is organized as follows. Inequalities involving mean opacities are discussed in section \ref{sec1}. The validity of Benford's law for line-strength collections is discussed in section \ref{sec2}. The observation made by Learner a long time ago and recently explained, that the number of lines of neutral iron are distributed in a specific way underlying a possible fractal structure of atomic spectra, is explained in section \ref{sec3}. The precision of interpolation is investigated in section \ref{sec4} and the convergence with respect to the maximum number of superconfigurations for given density and temperature is discussed in section \ref{sec5}.

\section{Opacity bounds}\label{sec1}

\subsection{The Rosseland mean opacity}\label{subsec12}

\begin{figure}[h]
\begin{center}
\includegraphics[width=9cm]{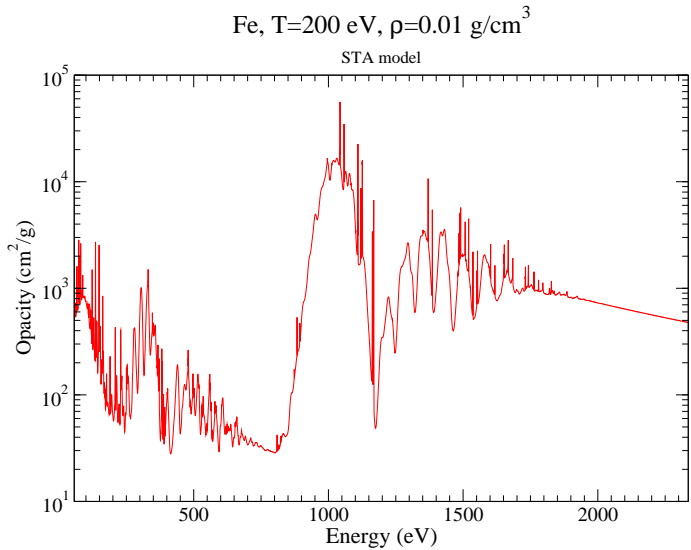} 
\end{center}
\caption{Opacity of an iron plasma at $T=$200 eV and $\rho$=0.01 g/cm$^3$.}\label{fig1}
\end{figure}

The Rosseland mean opacity is defined as \cite{Ribicki1979}:
\begin{equation}\label{defros}
\frac{1}{\kappa_R}=\frac{\int_0^{\infty}\displaystyle\frac{1}{\kappa(\nu)}\displaystyle\frac{dB(\nu)}{dT}d\nu}{\int_0^{\infty}\displaystyle\frac{dB(\nu)}{dT}d\nu},
\end{equation}
where $\kappa$ represents the radiative opacity and $T$ the temperature. $h$ denotes the Planck constant and $B(\nu)$ is the Planckian distribution
\begin{equation}
B(\nu)=\frac{2h\nu^3}{c^2}\left[\exp\left(\frac{h\nu}{k_BT}\right)-1\right]^{-1}.
\end{equation}
where $c$ is the speed of light in vacuum and $k_B$ the Boltzmann constant. Setting $u=h\nu/(k_BT)$, Eq. (\ref{defros}) becomes
\begin{equation}
\frac{1}{\kappa_R}=\int_0^{\infty}\frac{R(u)}{\kappa(u)}~du 
\end{equation}
with
\begin{equation}
R(u)=\frac{15}{4\pi^4}\frac{u^4e^u}{(e^u-1)^2}
\end{equation}
or explicitly
\begin{equation}
\kappa_R=\left(\frac{15}{4\pi^4}\int_0^{\infty}\frac{u^4e^{-u}}{\left(1-e^{-u}\right)^2}\frac{1}{\kappa(u)}~du\right)^{-1}.
\end{equation}
Figure \ref{fig1} represents the opacity of an iron plasma at $T=$200 eV and $\rho$=0.01 g/cm$^3$. It was computed by a code relying on the Super Transition Arrays formalism \cite{Barshalom1989,Salzmann1998,Pain2021,Pain2022}.

\subsection{From the Schwarz inequality to the Bernstein and Dyson bound}\label{subsec13}

Using the Schwarz inequality ($f$ and $g$ being functions of $u$) \cite{Schwarz1885}:
\begin{equation}
\left(\int fg~du\right)^2\leq\left(\int f^2~du\right)\left(\int g^2~du\right)
\end{equation}
with $f=\sqrt{R(u)/\kappa(u)}$ and $g=\sqrt{\kappa(u)}$, one gets
\begin{equation}
\left(\int_0^{\infty}\sqrt{R(u)}~du\right)^2\leq \mathscr{S}\frac{1}{\kappa_R} 
\end{equation}
with
\begin{equation}\label{integral}
\mathscr{S}=\int_0^{\infty}\kappa(u)~du=\frac{\pi e^2hZ\mathcal{N}_A}{4\pi\epsilon_0mcAk_BT}.
\end{equation}
The latter expression comes from the well-known Thomas-Reiche-Kuhn oscillator-strength sum rule \cite{Bethe,Merzbacher}. $Z$ is the atomic number, $\mathcal{N}_A$ the Avogadro number, $m$ the electron mass and $A$ the atomic mass. This leads to
\begin{equation}
\kappa_r\leq\frac{\mathscr{S}}{s^2}=\frac{\pi e^2hZ\mathcal{N}_A}{4\pi\epsilon_0mcAk_BTs^2},
\end{equation}
where $\zeta(3)$ is the Ap\'ery constant \cite{Vepstas2012}:
\begin{equation}
\zeta(3)=\frac{37\pi^3}{900}-\frac{2}{5}\sum_{n=1}^{\infty}\frac{1}{n^3}\left[\frac{4}{e^{\pi n}-1}+\frac{1}{e^{4\pi n}-1}\right],
\end{equation}
and $s=7\sqrt{15}~\zeta(3)/\pi^2\approx 3.30194$. One has approximately
\begin{equation}\label{bd}
\kappa_r\leq\frac{Z}{A}\frac{Ryd}{k_BT}\times 4.43\times 10^5 \mathrm{cm}^2/\mathrm{g}.
\end{equation}
The constraint in Eq. (\ref{integral}) can be easily improved by noting that the f-sum rule applies to the actual number of bound electrons in a given subshell, the atomic number $Z$ being replaced by the average number of bound electrons in the latter subshell at the specific density and temperature.

In the framework of the analysis of the pioneering Z-pinch experiments performed at Sandia \cite{Bailey2015} (showing an important discrepancy between experiment and theory) Iglesias pointed out that the measurements appear to violate the sum rule, but his analysis relies on a comparison with the ``cold opacity'', leading him to conclude that, since the main absorption features from the L shell are in the experimental range, the number of electrons in it would be inconsistent with the mean ionization (and even potentially larger than the degeneracy!) \cite{Iglesias2015}. This may be true, but is questionable because the Thomas-Kuhn-Reiche is valid for isolated-atom oscillator strengths and does not account for plasma effects and line shapes.

\subsection{Relation between Planck and Rosseland means}\label{subsec14}

The Planck mean opacity reads\footnote{The total frequency-dependent opacity can be calculated as the sum of the contributions of different processes: photo-excitation (or bound-bound opacity) $\kappa_{bb}$, photo-ionization (or bound-free opacity) $\kappa_{bf}$, inverse Bremsstrahlung $\kappa_{\mathrm{ff}}$ (or free-free opacity) and photon scattering $\kappa_{\mathrm{scat}}$. It is then given by the following expression: $\kappa(h\nu)=(\kappa_{\mathrm{bb}}(h\nu)+\kappa_{\mathrm{bf}}(h\nu)+\kappa_{\mathrm{ff}}(h\nu))(1-e^{-h\nu/k_BT})+\kappa_{\mathrm{scat}}(h\nu)$. However, in some definitions of the Planck mean opacity in connection with radiation-transfer modeling, the scattering contribution is not included \cite{Mihalas1978}. For simplicity here, we follow the work of Bernstein and Dyson \cite{Bernstein2003} and include the scattering contribution both in the Planck and Rosseland mean opacities. Since that contribution is usually much smaller than the others (except at very high frequency), and since we are looking for bounds, such an approximation seems reasonable. We also note that a factor $1-e^{-u}$ is missing in the denominator of the expression of $W_P(u)$ in Eq. (24) of Ref. \cite{Bernstein2003}.}
\begin{equation}
\kappa_P=\int_0^{\infty}P(u)\kappa(u)~du
\end{equation}
with
\begin{equation}
P(u)=\frac{15}{\pi^4}\frac{u^3e^{-u}}{1-e^{-u}}
\end{equation}
and thus we have
\begin{equation}\label{ratio}
\frac{\kappa_P}{\kappa_R}=\left(\int_0^{\infty}P(u)\kappa(u)~du\right)\left(\int_0^ {\infty}\frac{R(u)}{\kappa(u)}~du\right).
\end{equation}
Setting this time $f=\sqrt{P(u)\kappa(u)}$ and $g=\sqrt{R(u)/\kappa(u)}$, one gets
\begin{equation}
\frac{\kappa_P}{\kappa_R}\geq\left(\int_0^{\infty}\sqrt{P(u)R(u)}~du\right)^2\approx 0.949229
\end{equation}
yielding
\begin{equation}\label{arm}
\kappa_R\leq 1.05349~\kappa_P,
\end{equation}
which was obtained by Armstrong \cite{Armstrong1962}. However, since the Planck mean is often significantly larger than the Rosseland mean, such a relation is not really constraining.

\subsection{H\"older inequality}\label{subsec15}

The H\"older inequality reads \cite{Holder1889}
\begin{equation}
\left(\int fg~du\right)\leq \left(\int f^p~du\right)^{1/p}\left(\int g^q~du\right)^{1/q} 
\end{equation}
with 
\begin{equation}
\frac{1}{p}+\frac{1}{q}=1.
\end{equation}
Setting $f=\kappa(u)^{1/p}$ and $g=R(u)^{1/p}/\kappa(u)^{1/p}$, one gets
\begin{equation}\label{defK}
\mathscr{K}_p=\left(\int_0^{\infty}\frac{R(u)^{\frac{1}{p-1}}}{\kappa(u)^{\frac{1}{p-1}}}~du\right)\geq\frac{1}{\mathscr{S}}\left(\int_0^{\infty}R(u)^{1/p}~du\right)^p
\end{equation}
yielding
\begin{equation}
\mathscr{S}\geq\mathscr{S}_{\mathrm{min}}=\frac{1}{\mathscr{K}_p}\left(\int_0^{\infty}R(u)^{1/p}~du\right)^p.
\end{equation}
For $p=2$, the Schwarz inequality is recovered. The values of integrals $\left(\int_0^{\infty}[R(u)]^{p}~du\right)^{1/p}$ are displayed in table \ref{tab1} for $p$=2, 3, 4 and 5 together with the values of $\mathscr{K}_p$ (for an iron plasma at $T$=200 eV and $\rho$=0.25 g/cm$^3$). The lower bound for $\mathscr{S}$, still in the case of an iron plasma at $T$=200 eV and $\rho$=0.25 g/cm$^3$, is provided in table \ref{tab2}.
\begin{table}[]
\caption{Values of integrals $\left(\int_0^{\infty}[R(u)]^{1/p}~du\right)^p$ and $\mathscr{K}(p)$ (defined in Eq. (\ref{defK})) for $p$=2, 3, 4 and 5. $\mathscr{K}(p)$ is computed for an iron plasma at $T$=200 eV and $\rho$=0.25 g/cm$^3$.}
\centering
\begin{tabular}{c|c|c}
 $p$ & $\left(\int_0^{\infty}[R(u)]^{1/p}~du\right)^p$ & $\mathscr{K}(p)$ \\ \hline\hline
 2 & $\displaystyle\frac{735}{\pi^4}\left[\zeta(3)\right]^2\approx 10.90$ & $\approx 1.28~10^{-3}$\\ \hline
 3 & $\approx 158.96$ & $\approx 9.94~10^{-2}$\\ \hline
 4 & $\displaystyle\frac{15\pi^2\left[\Gamma\left(\frac{1}{4}\right)\right]^4}{4\left[\Gamma\left(\frac{3}{4}\right)\right]^4}\approx 2836.09$ & $\approx 5.24~10^{-1}$\\\hline
 5 & $\approx 59327.7$ & $\approx 1.29$\\ 
\end{tabular}
\label{tab1}
\end{table}

As can be checked in table \ref{tab2}, the lower bound for $\mathscr{S}$ is weaker (in the sense of ``less constraining'') than the Schwarz inequality ($p$=2) for $p$=3 and $p$=4, and stronger (\emph{i.e.} more constraining) in the case $p$=5.

\subsection{Introduction of an alternative mean opacity}\label{subsec16}

Setting, in the Schwarz inequality $f=\sqrt{R(u)/\kappa(u)}$ and $g=\sqrt{R(u)\kappa(u)}$, one gets, \emph{mutatis mutandis}
\begin{eqnarray}
& &\left(\int_0^{\infty}R(u)~du\right)^2\leq\left(\int_0^{\infty}\frac{R(u)}{\kappa(u)}~du\right)\nonumber\\
& &\times\left(\int_0^{\infty}R(u)\kappa(u)~du\right),
\end{eqnarray}
\emph{i.e.}
\begin{equation}
\kappa_M\geq\kappa_R 
\end{equation}
where we define the ``Milne opacity'' as
\begin{eqnarray}
\kappa_M&=&\int_0^{\infty}R(u)\kappa(u)~du\nonumber\\
&=&\frac{15}{4\pi^4}\int_0^{\infty}\frac{u^4e^{-u}}{\left(1-e^{-u}\right)^2}\kappa(u)~du. 
\end{eqnarray}
The latter quantity will be useful in the following, in order to test inequalities for the Rosseland mean.

\begin{table}[]
\caption{Values of the lower bound for $\mathscr{S}$ in the case of an iron plasma at $T$=200 eV and $\rho$=0.25 g/cm$^3$. The required values of $\left(\int_0^{\infty}[R(u)]^{1/p}~du\right)^{p}$ are provided in table \ref{tab1}.}
\centering
\begin{tabular}{c|c}
 $p$ & $\mathscr{S}_{\mathrm{min}}$\\ \hline\hline
 2 & $8513.39$\\ \hline
 3 & $1598.14$\\ \hline
 4 & $5412.19$ \\\hline
 5 & $45931.40$\\
\end{tabular}
\label{tab2}
\end{table}
\subsection{Milne inequalities for mixtures}\label{subsec17}

A long time ago, Milne used two interesting inequalities \cite{Milne1925}. The first one reads
\begin{equation}\label{milne1}
\left(\int fg~du\right)^2\leq \left(\int\left[f^2+g^2\right]~du\right)\left(\int\frac{f^2g^2}{
f^2+g^2}~du\right)
\end{equation}
and the second one
\begin{eqnarray}\label{milne2}
& &\left(\int\left[f^2+g^2\right]~du\right)\left(\int\frac{f^2g^2}{
f^2+g^2}~du\right)\nonumber\\
& &\leq \left(\int f^2~du\right)\left(\int g^2~du\right).
\end{eqnarray}
Starting from the second Milne inequality (see Eq. (\ref{milne2})), in the case of a mixture of two components, one has, setting $f^2=x_1R(u)/\kappa_1(u)$ and $g^2=x_2R(u)/\kappa_2(u)$:
\begin{equation}
\kappa_R\geq x_1\kappa_{R,1}+x_2\kappa_{R,2}.
\end{equation}
Applying the Schwarz inequality with $f^2=R(u)(x_1\kappa_1(u)+x_2\kappa_2(u))=R(u)\kappa(u)$ and $g^2=R(u)/\kappa(u)$, we get
\begin{equation}
\kappa_M=x_1\kappa_{M,1}+x_2\kappa_{M,2}\geq \kappa_R.
\end{equation}
Such a result can be generalized to the case of $n$ constituents (in other words a mixture of $n$ chemical elements) and one has
\begin{equation}\label{fulmilne}
\kappa_M=\sum_{i=1}^nx_i\kappa_{M,i}\geq \kappa_R\geq \sum_{i=1}^nx_i\kappa_{R,i}.
\end{equation}

\begin{figure}[h]
\begin{center}
\includegraphics[width=9cm]{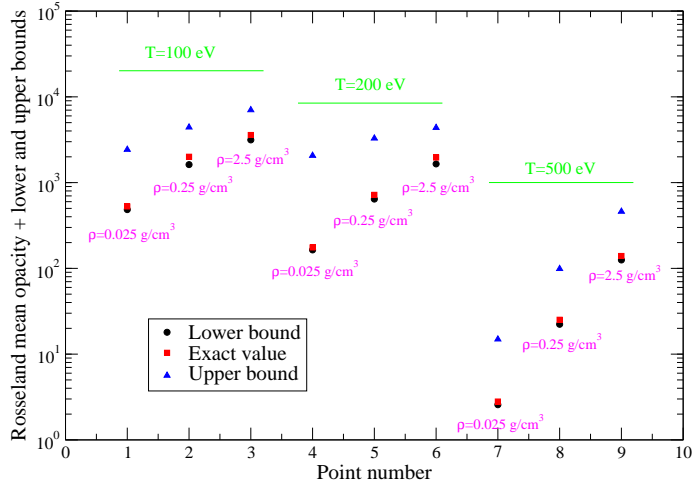}
\end{center}
\caption{Comparison between the exact Rosseland mean and the two bounds provided by Eq. (\ref{fulmilne}) in the case of an Fe-Mg (iron-magnesium) mixture plasma at three different temperatures: 100, 200 and 500 eV, and three different densities: 0.025, 0.25 and 2.5 g/cm$^3$.}\label{fig2}
\end{figure}

\begin{figure}[h]
\begin{center}
\includegraphics[width=9cm]{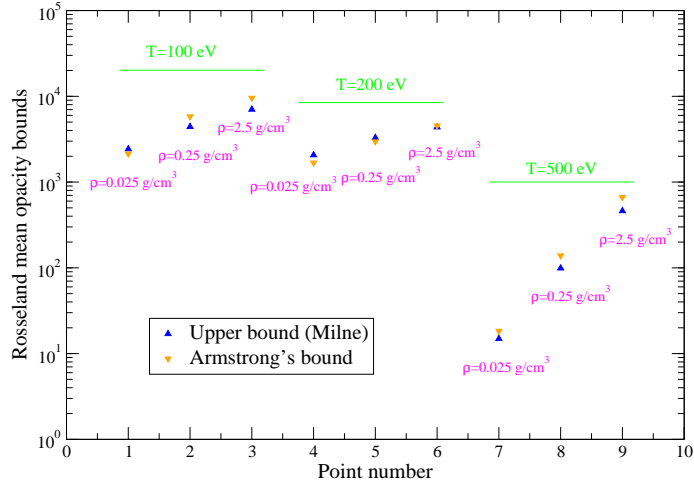}
\end{center}
\caption{Comparison between the upper bound of Eq. (\ref{fulmilne}) and Armstrong's bound \cite{Armstrong1962} (see Eq. (\ref{arm})) in the case of an Fe-Mg (iron-magnesium) mixture plasma at three different temperatures: 100, 200 and 500 eV, and three different densities: 0.025, 0.25 and 2.5 g/cm$^3$.}\label{fig2b}
\end{figure}

\subsection{A new bound using Milne identity}\label{subsec19}

Setting $f=\sqrt{\kappa(u)}$ and $g=\sqrt{R(u)/\kappa(u)}$ in the second inequality (\ref{milne2}), we get
\begin{eqnarray}
& &\frac{1}{\left(\int_0^{\infty}\left[\kappa(u)+\displaystyle\frac{R(u)}{\kappa(u)}\right]~du\right)}\nonumber\\
& &\times\frac{1}{\left(\int_0^{\infty}\displaystyle\frac{R(u)}{\kappa(u)+\displaystyle\frac{R(u)}{\kappa(u)}~du}\right)}\geq\frac{\kappa_R}{\mathscr{S}}\nonumber\\
& &
\end{eqnarray}
which looks as the first step of a continued fraction, yielding
\begin{equation}
\kappa_R\leq\frac{\mathscr{S}}{\left(\mathscr{S}+\displaystyle\frac{1}{\kappa_R}\right)\left(\int_0^{\infty}\displaystyle\frac{R(u)}{\kappa(u)+\displaystyle\frac{R(u)}{\kappa(u)}}~du\right)},
\end{equation}
which is more constraining that the Bernstein-Dyson bound (\ref{bd}). We have also
\begin{equation}
\int_0^{\infty}\displaystyle\frac{R(u)}{\kappa(u)+\displaystyle\frac{R(u)}{\kappa(u)}}~du\leq \int_0^{\infty}\displaystyle\frac{R(u)}{\kappa(u)}~du
\end{equation}
as well as

\begin{equation}
\int_0^{\infty}\frac{R(u)}{\kappa(u)}~du=\frac{1}{\kappa_R}\leq \int_0^{\infty}\frac{R(u)}{\kappa_{\mathrm{scatt}}(u)+\kappa_{\mathrm{IB}}(u)}~du
\end{equation}
and
\begin{eqnarray}
& &\int_0^{\infty}\frac{R(u)}{\kappa_{\mathrm{scatt}}(u)+\kappa_{\mathrm{IB}}(u)}~du\nonumber\\
&\leq&\min\left\{\int_0^{\infty}\frac{R(u)}{\kappa_{\mathrm{scatt}}(u)}~du,\int_0^{\infty}\frac{R(u)}{\kappa_{\mathrm{IB}}(u)}~du\right\}\nonumber\\
&=&\min\left\{\frac{1}{\kappa_{R,\mathrm{scatt}}},\frac{1}{\kappa_{R,\mathrm{IB}}}\right\},
\end{eqnarray}
where $\kappa_{\mathrm{scatt}}$ represents the scattering opacity in $\mathrm{cm}^2/\mathrm{g}$. It can be calculated in the Thomson approximation (no change in photon energy):
\begin{eqnarray}
\kappa_{\mathrm{scatt}}&=&\kappa_{R,\mathrm{scatt}}=\frac{8\pi}{3}\left(\frac{e^2}{4\pi\epsilon_0mc^2}\right)^2\frac{10^4~Z^*}{A[\mathrm{g}]}\mathcal{N}_A\nonumber\\
&\approx &0.665~10^{-24}\frac{Z^*\mathcal{N}_A}{A[\mathrm{g}]}
\end{eqnarray}
or more precisely (for high photon energies) using the Klein-Nishina formula (see Appendix). $\kappa_{\mathrm{IB}}$ the inverse-Bremsstrahlung contribution, calculated using the Kramers approximation \cite{Kramers1923} ($e$ represents the electron charge, $\epsilon_0$ the permittivity of vacuum and $Z^*$ the mean ionization of the plasma):
\begin{eqnarray}
\kappa_{\mathrm{IB}}(u)&=&\frac{16\pi^2}{3\sqrt{3}}\left(\frac{e^2}{4\pi\epsilon_0}\right)^3\frac{h^2}{(2\pi m)^{3/2}c}\frac{Z^{*3}}{(k_BT[\mathrm{eV}])^{7/2}u^3}\nonumber\\
& &\times\frac{\mathcal{N}_A^2}{(A[\mathrm{g}])^2}\left(\frac{10^{-3}}{e}\right)^{7/2}\rho[\mathrm{g}/\mathrm{cm}^3]\nonumber\\
&=&87.9~10^9~\frac{Z^{*3}\rho[\mathrm{g}/\mathrm{cm}^3]}{(A[\mathrm{g}])^2\left(h\nu[\mathrm{eV}]\right)^3\left(T[\mathrm{eV}]\right)^{1/2}}.
\end{eqnarray}
Using
\begin{equation}
\frac{15}{4\pi^4}\int_0^{\infty}\frac{u^7e^u}{(e^u-1)^2(1-e^{-u})}=\frac{10\left[\pi^6+945~\zeta(7)\right]}{\pi^4}
\end{equation}
with
\begin{equation}
\zeta(7)=\frac{409\pi^7}{94500}-\frac{2}{5}\sum_{n=1}^{\infty}\frac{1}{n^7}\left[\frac{4}{e^{\pi n}-1}+\frac{1}{e^{4\pi n}-1}\right],
\end{equation}
one finds
\begin{equation}
\kappa_{R,IB}[\mathrm{cm}^2/\mathrm{g}]=4.47283~10^8 \frac{Z^{*3}\rho[\mathrm{g}/\mathrm{cm}^3]}{(A[\mathrm{g})^2]\left(h\nu[\mathrm{eV}]\right)^3\left(T[\mathrm{eV}]\right)^{1/2}}.
\end{equation}

It is worth mentioning that it should also be possible to derive other new bounds using the second Milne inequality (\ref{milne2})\footnote{Setting $f^2(u)g^2(u)=R(u)$ and $f^2(u)+g^2(u)=\kappa(u)$ in the second inequality (\ref{milne2}) yields
\begin{eqnarray}
\frac{1}{\kappa_R}\leq\frac{1}{4\mathscr{S}}\left(\int_0^{\infty} [\kappa(u)-\sqrt{\kappa^2(u)-4R(u)}]~du\right)\nonumber\\
\times\left(\int_0^{\infty}[\kappa(u)+\sqrt{\kappa^2(u)-4R(u)}]~du\right)
\end{eqnarray}
provided that the arguments of the square roots are positive... This implies in any case
\begin{equation}
\kappa_R\geq \frac{1}{2\mathscr{S}}.
\end{equation}
}. The less known P\'olya-Szeg\"o's inequality \cite{Polya1925,Hardy1934,Zhao2013}, which states that if $0\leq m_1\leq f(u)\leq M_1$ and $0\leq m_2\leq g(u)\leq M_2$, then 

\begin{eqnarray}
& &\left(\int_0^{\infty} f^2~du\right)\left(\int_0^{\infty} g^2~du\right)\nonumber\\
&\leq&\frac{1}{4}\left(\sqrt{\frac{M_1M_2}{m_1m_2}}+\sqrt{\frac{m_1m_2}{M_1M_2}}\right)^2\left(\int_0^{\infty}fg~du\right)^2.\nonumber\\
& &
\end{eqnarray}
could lead to new constraints, as well as the ones published by Karamata \cite{Karamata1948} or Young \cite{Young1912}.

\section{Benford and the law of anomalous numbers}\label{sec2}

In 1881, Newcomb noticed that the first pages of logarithm books were more used than the last ones \cite{Newcomb1881}. Such an observation led to the conjecture that ``the significant digits of many sets of naturally occurring data are not equiprobably distributed, but in a way that favors smaller significant digits''. For instance, the first significant digit (\emph{i.e.} the first digit which is non zero) will be 6 more frequently than 7 and the first three significant digits will be 439 more often than 462. The law is verified by many sets of data: electricity bills, street addresses, stock prices, house prices, population numbers, death rates, lengths of rivers, Fibonacci and Lucas numbers, \emph{etc.} \cite{Wlodarski1971,Washington1981}. It is used to detect tax fraud and fraud in elections \cite{Barabesi2021}. Like other general principles about natural data - for example the fact that many data sets are well approximated by a normal distribution - there are illustrative examples and explanations that cover many of the cases where Benford's law applies, though there are many other cases where Benford's law applies that resist simple explanations \cite{Berger2020}. Benford's law tends to be most accurate when values are distributed across multiple orders of magnitude, especially if the process generating the numbers is described by a power law (which is common in nature).

\begin{figure}[h]
\begin{center}
\includegraphics[width=9cm]{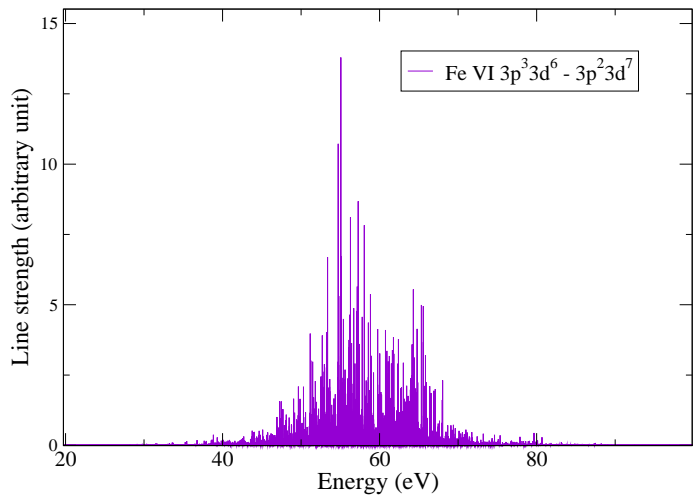}
\end{center}
\caption{Transition array $3p^33d^6-3p^23d^7$ of Fe VI (Fe$^{5+}$) computed with Cowan's code \cite{Cowan1981}.}\label{fig3}
\end{figure}

\begin{figure}[h]
\begin{center}
\includegraphics[width=9cm]{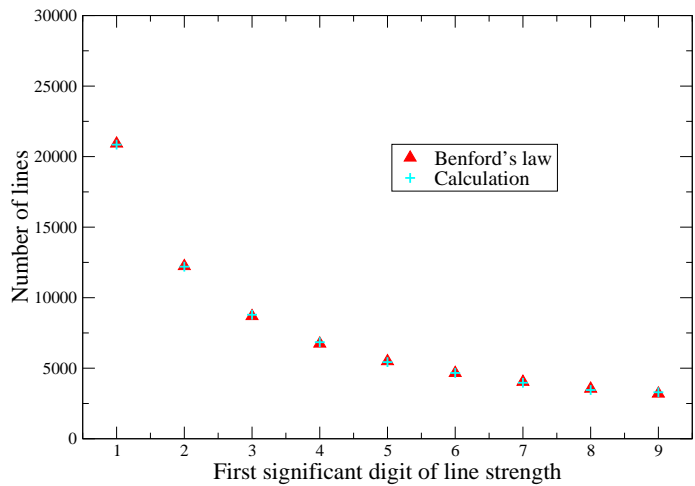}
\end{center}
\caption{Number of lines as a function of the first significant digit of their strength.}\label{fig4}
\end{figure}

In 1938, Benford found that the probability that the first significant digit $d_1$ is equal to $k$ is given by \cite{Benford1938}: 
\begin{equation}
\mathscr{P}\left(d_1=k\right)=\log_{10}\left(1+\frac{1}{k}\right).
\end{equation}
The law is valid for many kinds of data. The strength of a line $S_{ij}$ between levels $i$ and $j$ is defined, in atomic units, as
\begin{equation}
S_{ij}=\frac{3}{2}\frac{g_if_{ij}}{\Delta E_{ij}},
\end{equation}
where $g_i$ is the degeneracy of level $i$, $f_{ij}$ the oscillator strength of the line and $\Delta E_{ij}$ the line energy. We found recently that the distribution of line strengths in a given transition array follows very well Benford's logarithmic law of significant digits \cite{Pain2008,Pain2013}. The distribution of digits reflects the symmetries due to the selection rules; indeed, if transitions were governed by uncorrelated random processes, each digit would be equiprobable. It is worth mentioning that Benford's law is still not fully understood mathematically. Figure \ref{fig3} represents the transition array $3p^33d^6-3p^23d^7$ of Fe VI (Fe$^{5+}$) computed with Cowan's code \cite{Cowan1981} and Fig. \ref{fig4} shows a comparison between the actual distribution of the first digit and the prediction with Benford's law.

Since Benford's law can be explained in terms of a dynamics governed by multiplicative stochastic processes (additive in logarithmic space), Random Matrix Theory is an interesting tool for the calculation of large electric-dipole (E1) transition arrays \cite{Wilson1988}. 

\section{The Learner rule}\label{sec3}

Learner measured a large number of line intensities in the atomic spectrum of neutral iron and demonstrated in 1982 the existence of a remarkable power law for the density of lines versus their intensity \cite{Learner1982}: the logarithm 
\begin{equation}
\log_{10}\left(\mathscr{N}_n\right) 
\end{equation}
of the number of lines whose intensities lie between $2^nI_0$ and $2^{n+1}I_0$ ($n$ is an integer) is a decreasing linear function of $n$:
\begin{equation}
\log_{10}\frac{\mathscr{N}_n}{\mathscr{L}}\approx a_0-p\times n
\end{equation}
where $\mathscr{L}$ is the total number of lines, $a_0$ a constant and $-p$ the slope ($p$ is a positive real number). The value of $I_0$ is chosen in such a way that this rule holds for $1\leq n\leq$ 9 (9 octaves) when about 1500 lines within 290 nm $\leq\lambda\leq$ 550 nm are considered. One has
\begin{equation}
\mathscr{N}_n=\mathscr{N}_0.10^{-np},
\end{equation}
where $\mathscr{N}_0=10^{a_0}\mathscr{L}$ is the number of lines in the first octave: the number of lines is divided by $10^p$ when the size of the interval is multiplied by two. 

Learner observed that if $\mathscr{F}(k)$ is the number of lines having intensity in octave $k$: 
\begin{equation}\label{resuf}
\mathscr{F}(k)\approx \sqrt{2}~\mathscr{F}(k+1). 
\end{equation}
$\mathscr{F}(k)$ is computed through \cite{Bauche2015}
\begin{equation}
\mathscr{F}(k)=\int_{2^kI_0}^{2^{k+1}I_0}P(I)dI,
\end{equation}
$P(I)=\alpha I^{-3/2}$ being the intensity distribution. Equation (\ref{resuf}) is consistent with a distribution $P(I)=\alpha I^{-3/2}$. For fractal objects, the measured length may depend on the length of the measure:
\begin{equation}
\mathcal{L}(\ell)=\frac{K_0}{\ell^{D-1}},
\end{equation}
where $\mathcal{L}$ is the length of the object, $\ell$ the measure, $K_0$ a constant and $D$ the fractal dimension. In the present case, $\mathcal{L}$ can be chosen as the number of lines whose intensity is larger than $\ell$:
\begin{equation}
\mathcal{L}(\ell)=\int_{\ell}^{I_{\mathrm{max}}}\alpha~I^{-3/2}dI=2\alpha\left[\ell^{-1/2}-I_{\mathrm{max}}^{-1/2}\right]\approx 2\alpha\ell^{-1/2},
\end{equation} 
and neglecting $I_{\mathrm{max}}^{-1/2}$, one gets the fractal dimension $D=3/2$. For comparison, the dimension is $\ln 2/\ln 3\approx 0.63$ for the triadic Cantor set, $49\sqrt{3}/65\approx 1.31$ for the Apollonius circles, and $\ln 3/\ln 2\approx 1.58$ for the Sierpi\'nski triangle.
Recently, Fujii and Berengut reported that the combination of two statistical models - an exponential increase in the level density of many-electron atoms \cite{Dzuba2010} and local-thermodynamic-equilibrium excited-state populations - produces a surprisingly simple analytical explanation for this power-law dependence \cite{Fujii2019}. The authors found that the exponent of the power law is proportional to the electron temperature. This dependence may provide a useful diagnostic tool to extract the temperature of plasmas of complex atoms without the need to assign lines.

\section{Interpolations}\label{sec4}

Many efforts were devoted to the quantification of errors due to interpolations in opacity tables, as can be seen for instance in Ref. \cite{Seaton1993} for OP (Opacity Project) and Ref. \cite{Colgan2016} for ATOMIC opacities at Los Alamos. Similar remarks apply to OPAL data \cite{Iglesias1996,Rogers2002,Yang2008}. Independently, users of the tables can share their findings (see for instance Ref. \cite{Neuforge-Verheecke2001} in the framework of helioseismic tests of the new Los Alamos LEDCOP opacities).

We compare two grids covering wide ranges of temperatures and densities; the first one contains 2350 points (see Fig. \ref{fig11}) and the second one 21000 points (see Fig. \ref{fig12}).

\begin{figure}[h]
\begin{center}
\includegraphics[width=9cm]{grille_old_log.eps}
\end{center}
\caption{First grid containing 47 densities and 50 temperatures, \emph{i.e.} 2350 points.}\label{fig11}
\end{figure}

\begin{figure}[h]
\begin{center}
\includegraphics[width=9cm]{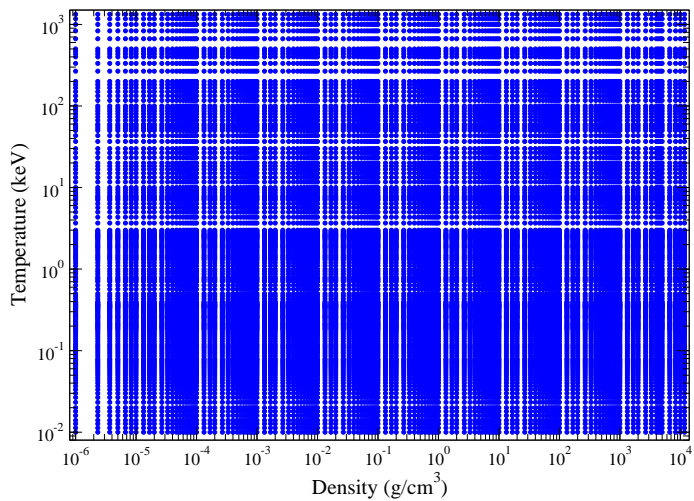}\label{fig12}
\end{center}
\caption{Second grid containing 140 densities and 150 temperatures, \emph{i.e.} 21000 points.}
\end{figure}

% xmgrace : S5 et S153 : 16 eV (celles du haut)

\begin{figure}[h]
\begin{center}
\includegraphics[width=9cm]{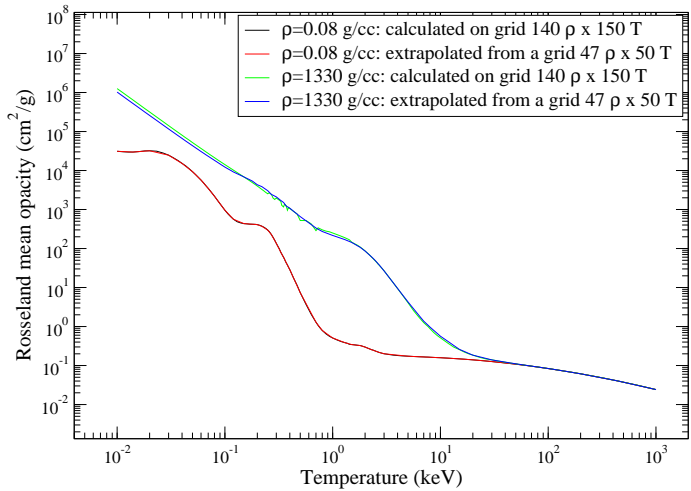}
\end{center}
\caption{Comparison between the iron Rosseland mean opacity calculated on the dense grid (140 densities and 150 temperatures), and interpolated on it from a smaller grid for two different densities: $\rho$=0.08 g/cm$^3$ and $\rho$=1330 g/cm$^3$. The latter isochores were chosen because they yield the highest discrepancies.}\label{fig13}
\end{figure}

Figure \ref{fig13} shows a comparison between the Rosseland mean opacity of iron calculated on the dense grid (140 densities and 150 temperatures), and interpolated on it from a smaller grid for two different densities: $\rho$=0.08 g/cm$^3$ and $\rho$=1330 g/cm$^3$. The latter isochores were chosen because they yield the highest discrepancies. Figure \ref{fig15} displays the same quantities but for two isotherms: $T$=16 eV and $T$=6.33 keV, which were chosen also because they are responsible for the most important discrepancies.

\begin{figure}[h]
\begin{center}
\includegraphics[width=9cm]{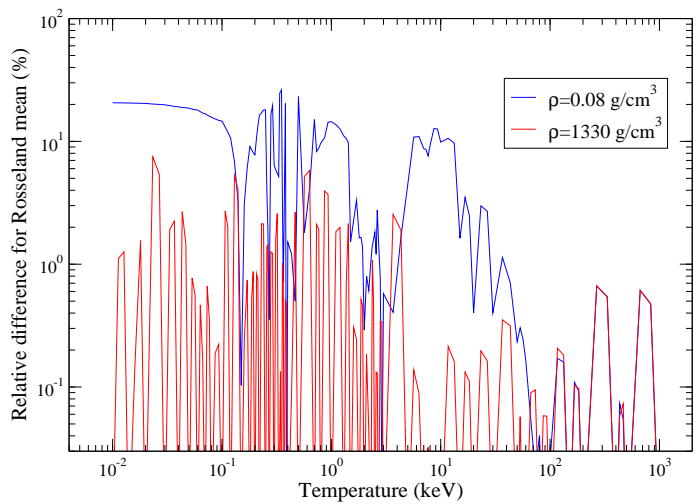}
\end{center}
\caption{Relative difference between the iron Rosseland mean opacity calculated on the extended grid (140 densities and 150 temperatures), and interpolated on it from a smaller grid for two different densities: $\rho$=0.08 g/cm$^3$ and $\rho$=1330 g/cm$^3$. The latter isochores were chosen because they yield the highest discrepancies.}\label{fig14}
\end{figure}

\begin{figure}[h]
\begin{center}
\includegraphics[width=9cm]{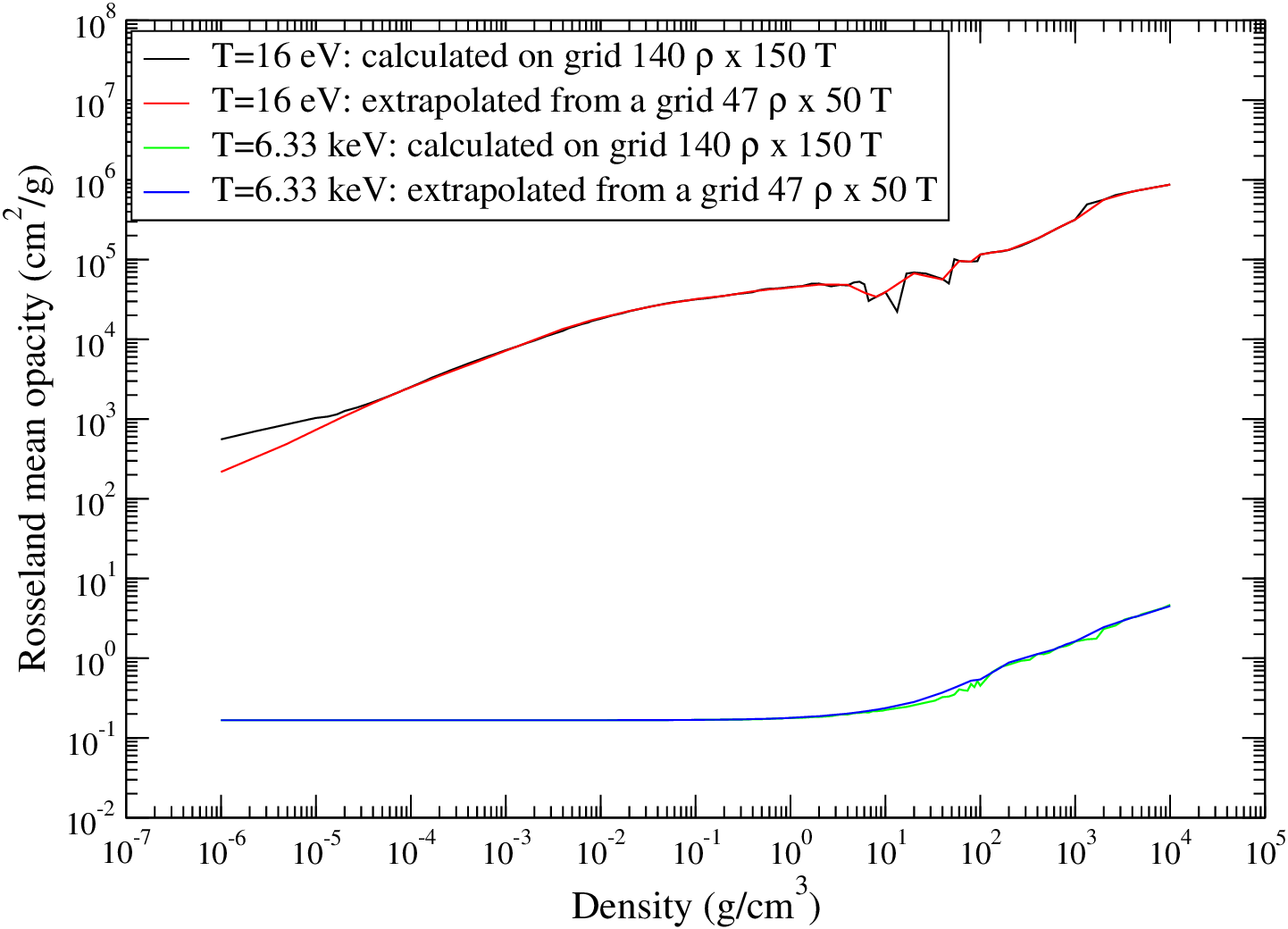}
\end{center}
\caption{Comparison between the iron Rosseland mean opacity calculated on the dense grid (140 densities and 150 temperatures), and interpolated on it from a smaller grid for two different densities: $T$=16 eV and $T$=6.33 keV. The latter isotherms were chosen because they yield the highest discrepancies.}\label{fig15}
\end{figure}

\begin{figure}[h]
\begin{center}
\includegraphics[width=9cm]{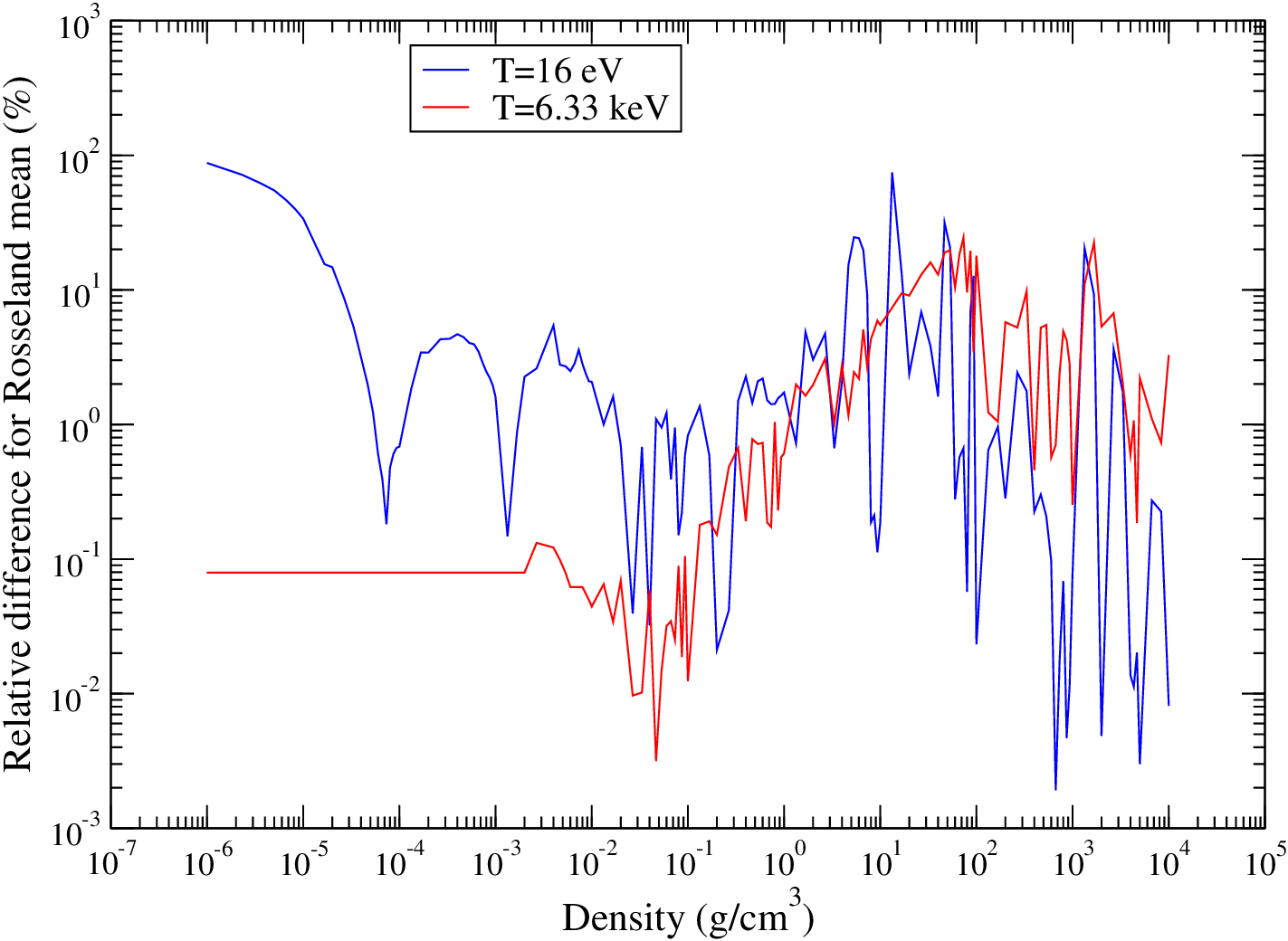}
\end{center}
\caption{Relative difference between the iron Rosseland mean opacity calculated on the dense grid (140 densities and 150 temperatures), and interpolated on it from a smaller grid for two different densities: $T$=16 eV and $T$=6.33 keV. The latter isotherms were chosen because they yield the highest discrepancies.}\label{fig16}
\end{figure}

Figures \ref{fig14} and \ref{fig16} display the relative difference between the iron Rosseland mean opacity calculated on the dense grid (140 densities and 150 temperatures), and interpolated on it from a smaller grid respectively for the two aforementioned densities (in the case of Fig. \ref{fig14}) and the two aforementioned temperatures (in the case of Fig. \ref{fig16}). 

Except in the latter cases, the differences are usually of the order of a few \% maximum. Of course, it is difficult to draw simple general conclusions from such an analysis, because we often have to perform simultaneously two interpolations: one for the density, and one for the temperature. The former can yield a small accuracy, while the latter not, and \emph{vice versa}.

\section{Convergence with respect to the number of superconfigurations}\label{sec5}

\begin{figure}[h]
\begin{center}
\includegraphics[width=9cm]{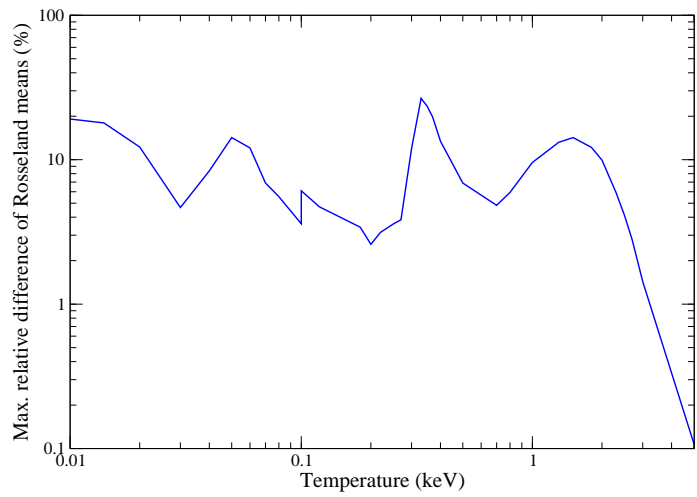}
\end{center}
\caption{Maximum value of the relative difference (in absolute values) between the iron Rosseland mean opacities of a calculation with a maximum number of 1000 superconfigurations and a maximum number of 10000 superconfigurations, as a function of temperature.}\label{fig6}
\end{figure}

\begin{figure}[h]
\begin{center}
\includegraphics[width=15.5cm]{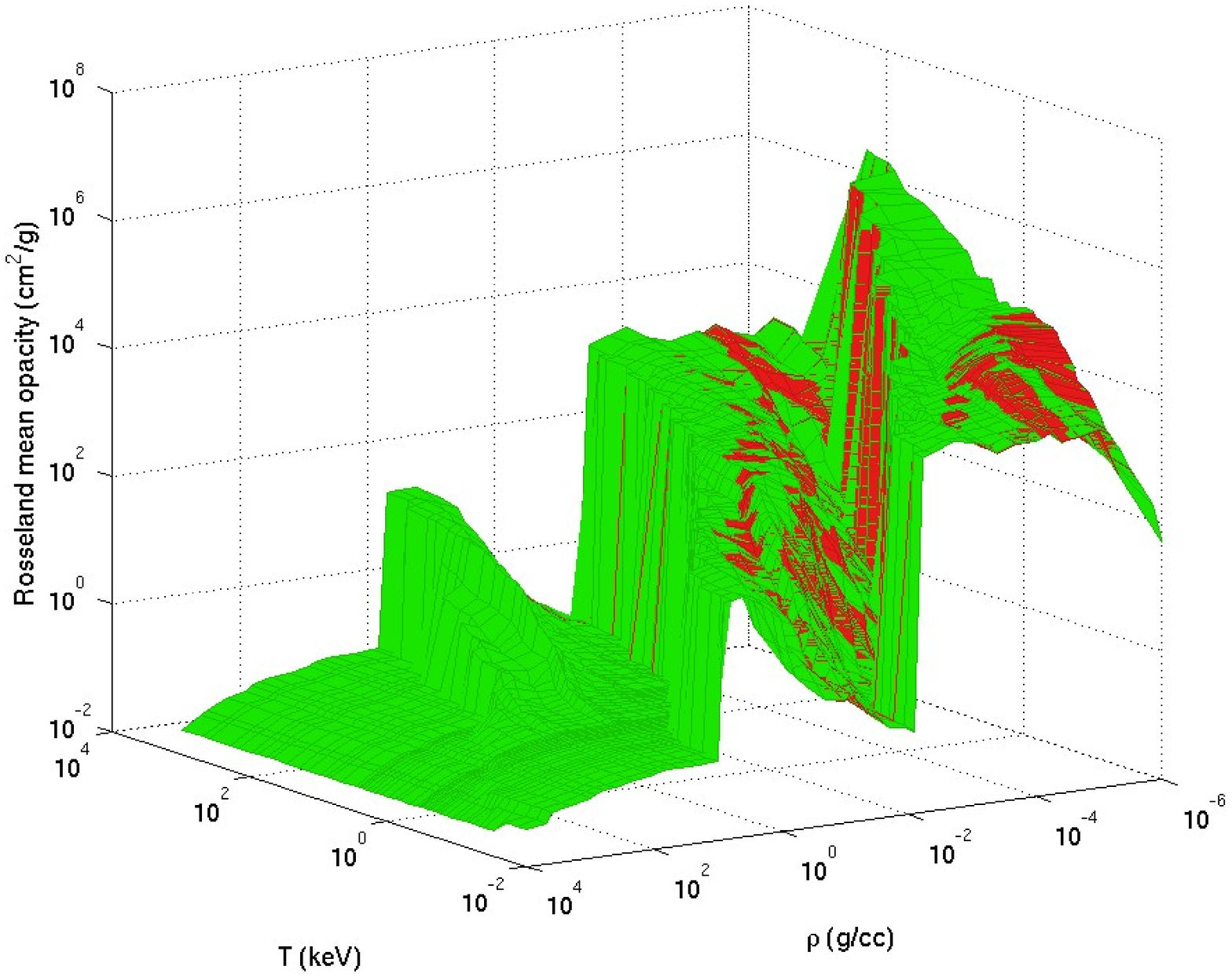}
\end{center}
\caption{Rosseland mean opacities of iron calculated with a maximum number of 1000 (blue curve) superconfigurations and a maximum number of 10000 superconfigurations (orange curve).}\label{fig9}
\end{figure}

\begin{figure}[h]
\begin{center}
\includegraphics[width=15.5cm]{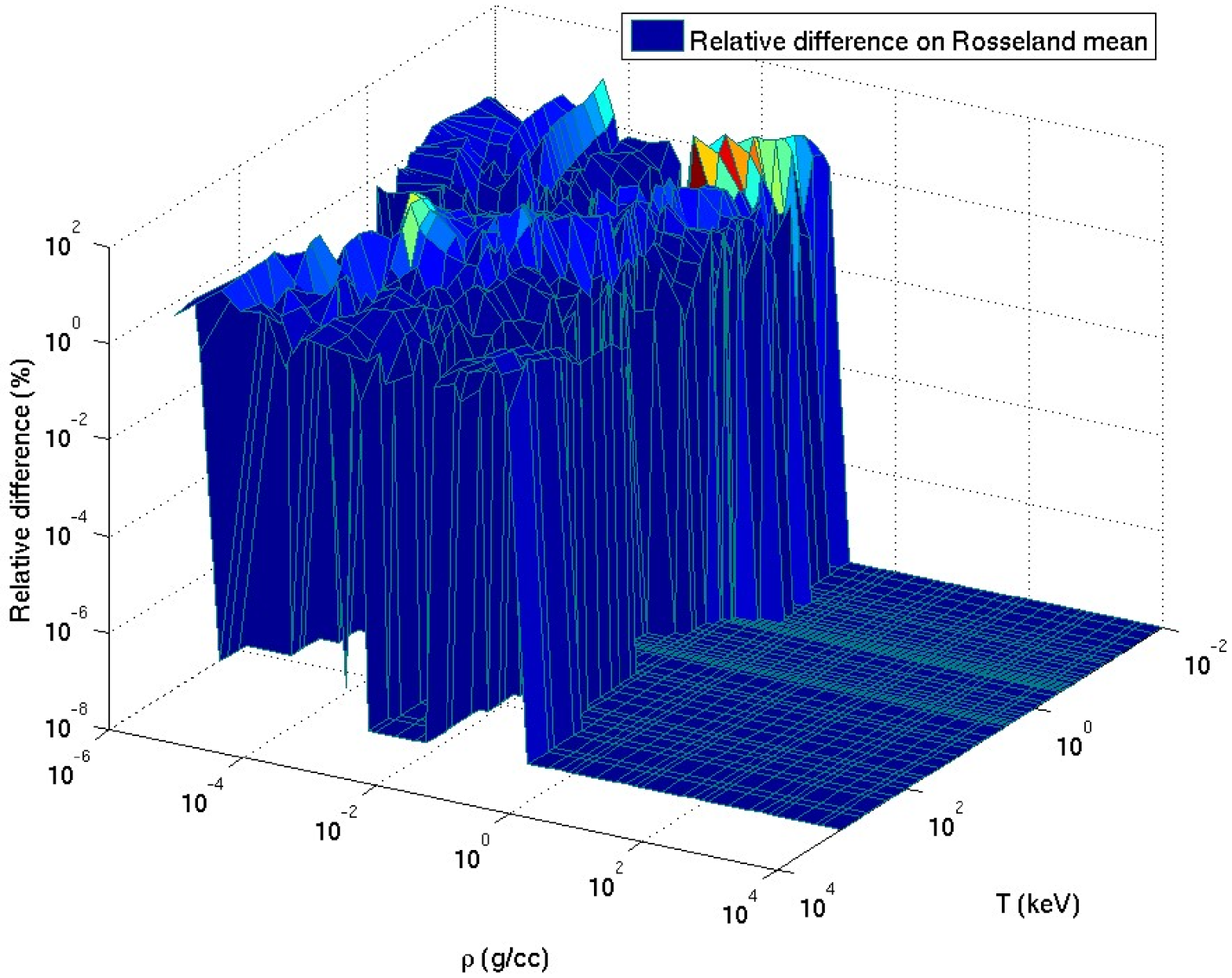}
\end{center}
\caption{Relative difference between the iron Rosseland mean opacities calculated with a maximum number of 1000 and 10000 superconfigurations.}\label{fig10}
\end{figure}

The STA model relies on the concept of superconfiguration. A superconfiguration is an ensemble of configurations close in energy. For instance,
\begin{equation}
S=(1s)^2(2s2p)^5(3s3p)^4(3d4s4p)^2(4d4f5s)^1
\end{equation}
is a superconfiguration made of five supershells: $(1s)$, $(2s2p)$, $(3s3p)$, $(3d4s4p)$ and $(4d4f5s)$ populated respectively with 2, 5, 4, 2 and 1 electron(s). For instance, $(3s3p)^4$ represents all the possibilities to distribute 4 electrons in $(3s)$ and $(3p)$, \emph{i.e.} the number of pairs $(a,b)$ such that $a+b=4$ and $0\leq a\leq 2$ and $0\leq b\leq 6$. The superconfiguration $S$ represents actually
\begin{equation}
\binom{8}{5}\times\binom{8}{4}\times\binom{18}{2}\times 26=15,593,760
\end{equation}
ordinary configurations, such as 
\begin{equation}
(1s)^2(2s)^2(2p)^3(3s)^1(3p)^3(3d)^1(4s)^1(4d)^1.
\end{equation}
We compare the cases with 1000 and 10000 superconfigurations. Such numbers are in fact the maximum numbers of superconfigurations for a density-temperature pair. In order to generate the list of superconfigurations, we use an adaptive ``divide and conquer'' algorithm, which ensures that we obtain, making successive gatherings and splittings of supershells, the optimum number of superconfigurations lower than the imposed maximum value \cite{Pain2022}. Figure \ref{fig6} represents the maximum value of the relative difference (in absolute values) between the iron Rosseland mean opacities of a calculation with a maximum number of 1000 superconfigurations and a maximum number of 10000 superconfigurations, as a function of temperature. We can see that the relative differences can reach 20 \% in some cases, which is very important. The iron Rosseland mean opacities computed with a maximum number of 1000 superconfigurations and a maximum number of 10000 superconfigurations are plotted in Fig. \ref{fig9}, and their relative difference in Fig. \ref{fig10}. The most important differences occur at high temperature and moderate density, when the number of excited states is important. A low density means a large Wigner-Seitz radius, and therefore more allowed subshells of high principal quantum number $n$ (and subsequently orbital angular momentum $\ell$), and a high temperature implies that high-lying states can be populated by electrons. However, as can be seen on the three aforementioned figures, things are a bit more complicated, this is just a general trend. 

\clearpage

\section{Conclusion}

We have presented different ideas and suggestions in order to limit the errors and uncertainties in opacity databases. Beyond the ones we mentioned, other mathematical inequalities may also be helpful \cite{Masjed2009}, such as the Minkowski inequality \cite{Minkowski1896,Woeginger2009}, or the Jensen convexity inequality \cite{Jensen1906}. It should also be very useful to resort to sum rules involving higher-order moments of the opacity, such as Eq. (\ref{molo}), as was proposed by Imshennik \emph{et al.} \cite{Imshennik1986} or Molodtsov \emph{et al.} \cite{Molodtsov1993}. One has to keep in mind, however, that such sum rules are extensions of ``pure atomic-physics'' ones, such as the Thomas-Reiche-Kuhn sum rule (also named ``oscillator-strength sum rule'' or ``$f-$sum rule''), which are valid for ``isolated atoms'' (where lines are Dirac delta functions in a sense), which makes their applicability questionable for hot and dense plasmas. In the same vein, it would be of great interest to try to derive similar bounds for molecular opacities \cite{Tennyson2018} using the mathematical tools mentioned in the present work. Actually, the sum rules for the opacity (like the simplest one $\int_0^{\infty}\kappa(u)~du$) will be different, since the opacity will not only contain an electronic part (similar as the one considered here), but also contributions of translation, rotation, and vibration, characteristic of molecules.

In addition, the constraints presented here can also be useful to assess the reliability of an experimental measurement \cite{Pain2020b}. If an experimentally inferred opacity (or any related quantity such as transmission, \emph{etc.}) does not fulfill one of the above mentioned inequalities, it means that something went wrong in the measurement. The bounds rely on two features: a mathematical inequality, and a sum rule. It is possible, however, that processes not yet in our knowledge could make correct experimental results appear to be incorrect. In addition to Iglesias' investigation on the Thomas-Reiche-Kuhn sum rule discussed above, it is worth mentioning that Liu et al. \cite{Liu2018} described a process that fails to satisfy the f-sum rule but agrees with experimental results.

We also pointed out the fact that Benford's law of anomalous numbers may enable one to detect errors in line-strength collections that are required in order to perform fine-structure calculations. Tables can also reveal regularities, such as the Learner rule, and bring information about the intrinsic properties of complex atomic spectra. Provided that they are confirmed, such regularities or trends can in turn help checking the relevance of tabulated data, for example through the calculation of a specific indicator as the fractal dimension. 

Finally, it is of course important to quantify the uncertainties due to interpolations in density-temperature opacity (or more generally atomic-data) tables and to ensure a proper convergence of the results.

\section*{Acknowledgments}

We would like to thank Alain Fontaine, Jean-Pierre Raucourt and Val\'erie Tabourin for their computational support and for helpful discussions. We are indebted to the anonymous referees for helpful comments and suggestions.

\section*{Appendix: Contribution of the Klein-Nishina scattering cross-section to the Rosseland mean}

We have seen that the Thomson opacity reads
\begin{equation}
\kappa_{\mathrm{Th}}=\frac{8\pi}{3}\left(\frac{e^2}{4\pi\epsilon_0mc^2}\right)^2\frac{Z^*}{A}\mathcal{N}_A.
\end{equation}
Let us introduce the reduced parameter
\begin{equation}
\gamma=\frac{h\nu}{mc^2}.
\end{equation}
The Klein-Nishina relativistic cross-section \cite{Klein1929} reads
\begin{eqnarray}
\kappa_{\mathrm{KN}}=&&\kappa_{\mathrm{Th}}\left\{\frac{1+\gamma}{\gamma^2}\left[\frac{2(1+\gamma)}{2\gamma+1}-\frac{1}{\gamma}\ln(2\gamma+1)\right]\right.\nonumber\\
& &\left.+\frac{1}{2\gamma}\ln(2\gamma+1)-\frac{3\gamma+1}{(2\gamma+1)^2}\right\}.
\end{eqnarray}
If $\gamma\ll 1$, one has
\begin{equation}\label{pg}
\kappa_{\mathrm{KN}}=\kappa_{\mathrm{Th}}\left(1-2\gamma+\frac{26}{5}\gamma^2+\cdots\right)
\end{equation}
and if $\gamma\gg1$:
\begin{equation}
\kappa_{\mathrm{KN}}\approx\frac{3}{8\gamma}\kappa_{\mathrm{Th}}\left[\ln(2\gamma)+\frac{1}{2}\right].
\end{equation}
Let us assume that the relativistic effects are negligible. We can then use Eq. (\ref{pg}) yielding
\begin{equation}
\frac{1}{\kappa_{\mathrm{KN}}}\approx\frac{1}{\kappa_{\mathrm{Th}}}\left(1+2\gamma-\frac{6}{5}\gamma^2\right).
\end{equation} 
Using the reduced variable $u=h\nu/(k_BT)$, one has for the Rosseland mean in the case of the Klein-Nishina expression of the scattering cross-section
\begin{eqnarray}
\kappa_{\mathrm{R,KN}}&=&\frac{4\pi^4}{15}\kappa_{\mathrm{Th}}\left[\int_0^{\infty}\frac{u^4e^{-u}}{(1-e^{-u})^2}\left(1+\frac{2k_BT}{mc^2}u\right.\right.\nonumber\\
& &\left.\left.-\frac{6}{5}\left(\frac{k_BT}{mc^2}\right)^2u^2\right)du\right]^{-1}
\end{eqnarray}
\emph{i.e.}
\begin{eqnarray}
\kappa_{\mathrm{R,KN}}&=&\frac{4\pi^4}{15}\kappa_{\mathrm{Th}}\left[\int_0^{\infty}\frac{u^4e^{-u}}{(1-e^{-u})^2}du\right.\nonumber\\
& &\left.+2\frac{k_BT}{mc^2}\int_0^{\infty}\frac{u^5e^{-u}}{(1-e^{-u})^2}du\right.\nonumber\\
& &\left.-\frac{6}{5}\left(\frac{k_BT}{mc^2}\right)^2\int_0^{\infty}\frac{u^6e^{-u}}{(1-e^{-u})^2}du\right]^{-1}.
\end{eqnarray}
Using
\begin{equation}
\int_0^{\infty}\frac{u^4e^{-u}}{(1-e^{-u})^2}du=\frac{4\pi^4}{15}
\end{equation}
as well as
\begin{equation}
\int_0^{\infty}\frac{u^5e^{-u}}{(1-e^{-u})^2}du=120~\zeta(5)
\end{equation}
and
\begin{equation}
\int_0^{\infty}\frac{u^6e^{-u}}{(1-e^{-u})^2}du=\frac{16\pi^6}{21},
\end{equation}
one finally obtains
\begin{equation}
\kappa_{\mathrm{R,KN}}=\kappa_{\mathrm{Th}}\left[1+\frac{900~\zeta(5)}{\pi^4}\left(\frac{k_BT}{mc^2}\right)-\frac{24\pi^2}{7}\left(\frac{k_BT}{mc^2}\right)^2\right]
\end{equation}
and using \cite{Plouffe1998,Chamberland2011}:
\begin{eqnarray}
\zeta(5)&=&\frac{\pi^5}{294}-\frac{72}{35}\sum_{n=1}^{\infty}\frac{1}{n^5\left(e^{2\pi n}-1\right)}-\frac{2}{35}\sum_{n=1}^{\infty}\frac{1}{n^5\left(e^{2\pi n}+1\right)}\nonumber\\
&\approx& 1.03692,
\end{eqnarray}
one has
\begin{equation}
\kappa_{\mathrm{R,KN}}\approx\kappa_{\mathrm{Th}}\left[1+9.58057\left(\frac{k_BT}{mc^2}\right)-33.8386\left(\frac{k_BT}{mc^2}\right)^2\right].
\end{equation}


\begin{thebibliography}{}

\bibitem{Armstrong1972} 
B. H. Armstrong and R. W. Nicholls, {\it Emission, Absorption and Transfer of Radiation in Heated Atmospheres} (Pergamon Press, Oxford, 1972).

\bibitem{Rybicki1985} G. B. Rybicki and A. P. Lightman, {\it Radiative processes in astrophysics} (Wiley, New York, 1985).

\bibitem{Huebner2016}
W. F. Huebner and W. D. Barfield, {\it Opacity} (Springer, New York, 2016).

\bibitem{Michaud2016}
G. Michaud, G. Alecian and J. Richer, {\it Atomic Diffusion in Stars} (Springer Cham, 2016).

\bibitem{Pain2018b} J.-C. Pain and F. Gilleron, Opacity calculations for stellar astrophysics Proceedings of the PHOST ``Physics of Oscillating Stars'' conference, 2-7 Sept. 2018, Banyuls-sur-mer (France).\\
\url{https://doi.org/10.5281/zenodo.1590773}\\
\url{https://zenodo.org/record/1590773#.YY59L06ZPDc}

\bibitem{Ralchenko} Y. Ralchenko, {\it Modern Methods in Collisional-Radiative Modeling of Plasmas} (Springer International Publishing AG; 1st ed. 2016).

\bibitem{Bauche2015} J. Bauche, C. Bauche-Arnoult and O. Peyrusse, {\it Atomic Properties in Hot Plasmas. From Levels to Superconfigurations} (Springer International Publishing, 2015).

\bibitem{Dyson2002} G. B. Dyson, {\it Project Orion} (New York: Henry Holt, 2002).

\bibitem{Bernstein1959} J. Bernstein and F. J. Dyson, The continuous opacity and equations of state of light elements at low densities (General Atomic Report GA-848, unpublished).

% Opacity bounds
\bibitem{Bernstein2003} J. Bernstein and F. Dyson, PASP \textbf{115}, (2003) 1383-1387.

% Maximum opacity theorem
\bibitem{Armstrong1962} B. H. Armstrong, Astrophys. J. {\bf 136}, (1962) 309-310.

% Lower bound on the Rosseland mean free path 
\bibitem{Imshennik1986} V. S. Imshennik, I. N. Mikhailov, M. M. Basko and S. V. Molodtsov, Sov. Phys. JETP \textbf{63}, (1986) 980-985.

% A complete system of estimates of the minimal Rosseland mean free path of photons on the basis of sum rules
\bibitem{Molodtsov1993} S. V. Molodtsov, J. Exp. Theor. Phys. \textbf{77}, (1993) 406-412. 

\bibitem{Ribicki1979} G. B. Ribicki and A. P. Lightman, {\it Radiative Processes in Astrophysics} (New York, Wiley, 1979).

% Super-transition-arrays: A model for the spectral analysis of hot, dense plasmas
\bibitem{Barshalom1989} A. Bar-Shalom, J. Oreg, W. H. Goldstein, D. Shvarts and A. Zigler, Phys. Rev. A {\bf 40}, (1989) 3183-3193.

\bibitem{Salzmann1998} D. Salzmann, {\it Atomic Physics in Hot Plasmas} (Oxford University Press, New York and Oxford, 1998).

% Super Transition Arrays: A tool for studying spectral properties of hot plasmas
\bibitem{Pain2021} J.-C. Pain, Plasma {\bf 3}, (2021) 42-64. 

% Adaptive algorithm for the generation of superconfigurations in hot-plasma opacity calculations
\bibitem{Pain2022} J.-C. Pain, Plasma {\bf 5}, (2022) 154-175.

% \"Uber ein Fl\"achen kleinsten Fl\"acheninhalts betreffendes Problem der Variationsrechnung
\bibitem{Schwarz1885} H. A. Schwarz, Acta Societatis Scientiarum Fennicae {\bf 15}, (1885) 315-362.

\bibitem{Bethe} H. Bethe and E. Salpeter, {\it Quantum Mechanics of One- and Two-electron Atoms} (Academic Press, New York, 1957).

\bibitem{Merzbacher} E. Merzbacher, {\it Quantum Mechanics} (Wiley, New York, 1970).

% On Plouffe's Ramanujan identities
\bibitem{Vepstas2012} L. Vep$\mathrm{\check{s}}$tas, Ramanujan J. {\bf 27}, (2012) 387-408.

% A higher-than-predicted measurement of iron opacity at solar interior temperatures
\bibitem{Bailey2015} 
Bailey, J. E.; Nagayama, T.; Loisel, G. P.; Rochau, G. A.; Blancard, C.; Colgan, J.; Coss\'e, P.; Faussurier, G.; Fontes, C. J.; Gilleron, F.; Golovkin, I.; Hansen S. B.; Iglesias, C. A.; Kilcrease, D. P.; MacFarlane, J. J.; Mancini, R. C.; Nahar, S. N.; Orban, C.; Pain, J.-C.; Pradhan, A. K., Sherrill, M.; Wilson, B. G. A higher-than-predicted measurement of iron opacity at solar interior temperatures. {\it Nature} {\bf 2015}, {\it 517}, 56-59.

% Enigmatic photon absorption in plasmas near solar interior conditions
\bibitem{Iglesias2015} 
C. A.Iglesias, {\it High Energy Density Phys.} {\bf 15}, (2015) 4-7.

\bibitem{Mihalas1978} D. Mihalas, {\it Stellar Atmospheres} (W. H. Freeman, 1978).

% Uber einen Mittelwerthsatz
\bibitem{Holder1889} O. L. H\"older, Nachrichten von der K\"oniglichen Gesellschaft der Wissenschaften und der Georg-Augusts-Universit\"at zu G\"ottingen (1889) 38-47.

% Note on Rosseland's integral for the stellar absorption coefficient
\bibitem{Milne1925} E. A. Milne, Proceedings of the Royal Astronomical Society \textbf{43}, (1925) 979-984.

% On the theory of X-ray absorption and of the continuous X-ray spectrum
\bibitem{Kramers1923} H. A. Kramers, Phil. Mag. {\bf 46}, (1923) 836-871. 

\bibitem{Polya1925} G. P\'olya and G. Szeg\"o, {\it Aufgaben aus der Analysis}, vol. I (Springer, Berlin, 1925).

\bibitem{Hardy1934} G. H. Hardy, J. E. Littlewood and G. P\'olya, {\it Inequalities} (Cambridge University Press, Cambridge, 1934).

% On P\'olya-Szeg\"o's inequality
\bibitem{Zhao2013} C.-J. Zhao and W.-S. Cheung, J. Inequalities Appl. {\bf 2013}, (2013) 591-595. 

% Sur certaines in\'egalit\'es relatives aux quotients et \`a la diff\'erence de $\int fg$ et $\int f\int g$
\bibitem{Karamata1948} J. Karamata, Acad. Serbe Sci. Publ. Inst. Math. {\bf 2}, (1948) 131-145.

% On classes of summable functions and their Fourier series
\bibitem{Young1912} W. H. Young, Proc. Roy. Soc. Lond. Series A {\bf 87}, (1912) 225-229.

% Note on the frequency of use of the different digits in natural numbers
\bibitem{Newcomb1881} S. Newcomb, Am. J. Math. {\bf 4}, (1881) 39-40.

% Fibonacci and Lucas Numbers tend to obey Benford's law
\bibitem{Wlodarski1971} J. Wlodarski, Fibonacci Q. {\bf 9}, (1971) 87-88.

% Benford's law for Fibonacci and Lucas numbers 
\bibitem{Washington1981} L. C. Washington, Fibonacci Q. {\bf 19}, (1981) 175-177. 

% Forum on Benford's law and statistical methods for the detection of frauds
\bibitem{Barabesi2021} L. Barabesi, A. Cerioli and D. Perrotta, Stat. Meth. Appl. {\bf 30}, (2021) 767-778.

% The mathematics of Benford's law: a primer
\bibitem{Berger2020} A. Berger and T. P. Hill, Stat. Methods Appl. {\bf 30}, (2020) 779-795.

% The law of anomalous numbers
\bibitem{Benford1938} F. Benford, Proc. Am. Philos. Soc. {\bf 78}, (1938) 551-572.

% Benford's law and complex atomic spectra
\bibitem{Pain2008} J.-C. Pain, Phys. Rev. E {\bf 77}, (2008) 012102.

% Regularities and symmetries in atomic structure and spectra
\bibitem{Pain2013} J.-C. Pain, High Energy Density Phys. {\bf 9}, (2013) 392-401.

\bibitem{Cowan1981} R. D. Cowan, {The theory of atomic structure and spectra} (University of California Press, 1981).

% Random-matrix method for the simulation of large atomic E1 transition arrays
\bibitem{Wilson1988} B. G. Wilson, F. Rogers and C. Iglesias, Phys. Rev. A {\bf 37}, (1988) 2695-2697.

% A simple (and unexpected) experimental law relating to the number of weak lines in a complex spectrum
\bibitem{Learner1982} R. C. M. Learner, J. Phys. B: Atom. Mol. Phys. {\bf 15}, (1982) L891-L895.

% Exponential increase of energy level density in atoms: Th and Th II
\bibitem{Dzuba2010} V. A. Dzuba et V. V. Flambaum, Phys. Rev. Lett. {\bf 104}, (2010) 213002. 

% A Simple Explanation for the Observed Power Law Distribution of Line Intensity in Complex Many-Electron Atoms
\bibitem{Fujii2019} K. Fujii and J.-C. Berengut, Phys. Rev. Lett. {\bf 124}, (2019) 185002. 

% Fitting and smoothing of opacity data
\bibitem{Seaton1993} M. J. Seaton, Mon. Notices Royal Astron. Soc. {\bf 265}, (1993) L25-L28.

% A New Generation of Los Alamos Opacity Tables
\bibitem{Colgan2016} 
J. Colgan, D. P. Kilcrease, N. H. Magee, M. E. Sherrill, J. Abdallah Jr., P. Hakel, C. J. Fontes, J. A. Guzik and K. A. Mussack, {\it Astrophys. J.} {\bf 817}, (2016) 116.

% Updated OPAL opacities
\bibitem{Iglesias1996} 
C. Iglesias and F. J. Rogers, {\it Astrophys. J.} {\bf 464}, (1996) 943-953.

% Updated and expanded OPAL equation-of-state tables: implications for helioseismology
\bibitem{Rogers2002} 
F. J. Rogers and A. Nayfonov, {\it Astrophys. J.} {\bf 576}, (2002) 1064-1074. 

% Asteroseismic constraints on the OPAL opacity interpolation
\bibitem{Yang2008} W. Yang and M. Li, Proc. In. Astron. Union, {\bf 4(S252)}, (2008) 123-124.

% Helioseismic tests of the new Los Alamos LEDCOP opacities
\bibitem{Neuforge-Verheecke2001} C. Neuforge-Verheecke, J. A. Guzik, J. J. Keady, N. H. Magee, P. A. Bradley and A. Noels, {\it Astrophys. J.} {\bf 561}, (2001) 450-454 .

% A functional generalization of the Cauchy-Schwarz inequality and some subclasses
\bibitem{Masjed2009} M. Masjed-Jamei, Appl. Math. Lett. {\bf 22}, (2009) 1335-1339.

\bibitem{Minkowski1896} H. Minkowski, {\it Geometrie der Zahlen} (Teubner, Leipzig, 1896).

% When Cauchy and H\"older met Minkowski: A tour through well-known inequalities
\bibitem{Woeginger2009} G. J. Woeginger, Math. Mag. {\bf 82}, (2009) 202-207.

% Sur les fonctions convexes et les in\'egalit\'es entre les valeurs moyennes
\bibitem{Jensen1906} J. L. W. V. Jensen, Acta Math. {\bf 30}, (1906) 175-193.

% The ExoMol Atlas of Molecular Opacities
\bibitem{Tennyson2018} J. Tennyson and S. N. Yurchenko, Atoms {\bf 6}, (2018) 26.

% A quantitative study of some sources of uncertainty in opacity measurements
\bibitem{Pain2020b} J.-C. Pain and F. Gilleron, High Energy Density Phys. {\bf 34}, (2020) 100745.

% Transient space localization of electrons ejected from continuum atomic processes in hot dense plasma
\bibitem{Liu2018} P. Liu, C. Gao, Y. Hou, J. Zeng and J. Yuan, Commun. Phys. {\bf 1}, 95 (2018).

% \"Uber die Streuung von Strahlung durch freie Elektronen nach der neuen relativistischen Quantendynamik von Dirac
\bibitem{Klein1929} O. Klein and Y. Nishina, Z. Phys. {\bf 52}, (1929) 853-868.

\bibitem{Plouffe1998} S. Plouffe, {\it Identities inspired from Ramanujan Notebooks II}, \url{http://www.plouffe.fr/simon/identities.html}\\
\url{http://www.plouffe.fr/simon/constants/zeta5.txt}

% Formulas for Odd Zeta Values and Powers of $\pi$
\bibitem{Chamberland2011} M. Chamberland and P. Lopatto, J. Integer Seq. {\bf 14}, (2011) article 11.2.5. 

\end{thebibliography}
\end{document}